# Anti-biofouling Lensless Camera System with Deep Learning based Image Reconstruction

Naoki Ide[1], Tomohiro Kawahara[1], Hiroshi Ueno[1], Daiki Yanagidaira[1], and Susumu Takatsuka[1, 2]

1 Advanced Research Laboratory, Sony Group Corporation, Tokyo, Japan
2 Super-cutting-edge Grand and Advanced Research (SUGAR) Program, Japan Agency for Marine-Earth Science and Technology (JAMSTEC), Yokosuka, Japan

*Abstract*—In recent years, there has been an increasing demand for underwater cameras that monitor the condition of offshore structures and check the number of individuals in aqua culture environments with long-period observation. One of the significant issues with this observation is that biofouling sticks to the aperture and lens densely and prevents cameras from capturing clear images. This study examines an underwater camera that applies material technologies with high inherent resistance to biofouling and computer vision technologies based on image reconstruction by deep learning to lens-less cameras. For this purpose, our prototype camera uses a coded aperture with 1k rectangular shape pinholes in a thin metal plate, such as copper, which hinder the growth of biofouling and keep the surface clean. Although images taken by lens-less cameras are usually not well formed due to lack of the traditional glass-based lens, a deep learning approach using ViT (Vision Transformer) has recently demonstrated reconstructing original photo images well and our study shows that using gated MLP (Multilayer Perceptron) also yields good results. On the other hand, a certain degree of thickness for bio-repellence materials is required to exhibit their effect the thickness of aperture is necessary to use apertures sufficiently thinner than the size of the pinholes to avoid unintentional reflection and absorption on the sidewalls. Therefore, we prepared a sufficiently thin plate for image reconstruction and now currently we conduct tests of the lens-less camera of the bio-repellence aperture with actual seawater environments to determine whether it can sufficiently demonstrate the biofouling effect compared with usual camera with only waterproof.

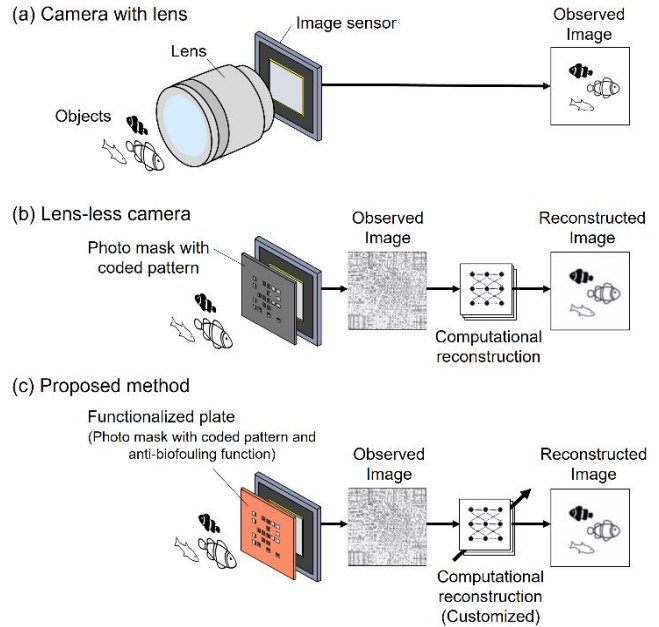

Figure 1: Comparison between camera-based sensing approaches.

## I. INTRODUCTION

The importance of underwater *in-situ* monitoring has been increasing obviously in many fields such as global environment observation, marine conservation, marine omics, fisheries industry, and so on [1, 2]. Even through camera device is suitable for long-time monitoring, it is highly difficult to apply the camera for long-term sensing due to the harsh condition of the underwater environment.

In the typical camera device for land applications, the components of the camera especially lens parts can be easily cleaned and maintained by users to obtain long-term imaging performance, as shown in Fig. 1(a). In contrast, if the camera is located in a remote and underwater environment, the maintenance process becomes much harder and the dirt accumulating on the lens will continue to expand until the quality of the image captured through the lens falls below acceptable standards. To clarify the mechanism of the dart accumulation by the biological adhesion, many works have been conducted and mechanical wiping and ultraviolet irradiation methods have been introduced to maintain the surface condition of the lens of the monitoring devices [3, 4]. However, the durability and the power consumption of these mechanisms are still challenging factors to realize a distributed camera system in the underwater environment. Even through applying the anti-fouling coatings to camera lenses is contribute to maintained the performance and will be a great candidate, there are limitations in terms of design flexibility, durability, and production cost.

On the other hand, in the field of the computational optics, lens-less cameras are gaining attention as a technology that significantly deviates from traditional imaging systems, as shown in Fig. 1(b). These cameras operate without using a lens and instead, they reconstruct the image of the subject through calculations based on digitized image data captured by an image sensor. In other words, computational algorithms replace the physical imaging function of the lens. This approach, which eliminates the lens in a camera, also removes various constraints associated with conventional imaging systems that rely on lens-based apertures. Furthermore, it is anticipated that the aperture can assume new roles beyond merely capturing light.

Based on these backgrounds, we newly propose a lens-less camera system which is composed of an image sensor and a photo mask (a functionalized plate) for the underwater monitoring applications over a long period of time, as shown in Fig. 1(c). In our approach, the functionalized plate is designed to realize multiple functions such as the image coding and the anti-fouling of the camera system. In addition, customized computational reconstruction techniques such as deep-leaning are actively introduced for realizing a fine lens-less imaging with a higher environmental resistance. In this paper, we present detail of the basic concept of the proposed method and show an example of how it can be implemented and performed .

After that, our unique contributions, particularly through the use of deep learning will be examine through the experiments. Finally, we will discuss the challenges associated with the proposed approach, insights gained from balancing dirt avoidance with lens-less image quality.

## II. RELATED WORKS

### A. Lens-less camera and imaging

The simplest camera without a lens is the pinhole camera. A mask with a single tiny hole instead of a lens is placed in front of the sensor to take pictures. However, this camera has the problem of producing a dark image due to the low light level. To solve this problem, sensing is performed using a plate with multiple tiny holes, called a coded aperture [5]. The picture acquired by the sensor is an out-of-focus, blurred image due to the multiple overlapping of the image incident through the multiple holes. If the position of the holes is invariant, the overlap of the images will have an invariant regularity, the original image can be reconstructed by deep learning inference by learning with a large amount of teacher data [6].

In particular, a high-resolution image reconstruction and a functionalization of lens-less imaging have been proposed and implemented by focusing the recent computing power [7]. However, there are currently no studies focusing on combining the functionality of coding elements with computation to address significant variations in real-world application such as marine environments.

### B. Anti-fouling approaches

In the field of marine technology, a wide range of coating technologies have traditionally been developed to enhance the durability of the bottom of ships and the surface of the off-shore facilities [8]. In these cases, anti-fouling coating paints containing metallic components are typically used and their durability can be continued for several months to a year level [9]. Even through non-biocide type materials are recently used to minimize impact on marine organisms [10], they have a limited anti-fouling performance.

There is still limited research on complementary system that actively combining the anti-biofouling functionality with computational approach have not been sufficiently conducted yet.

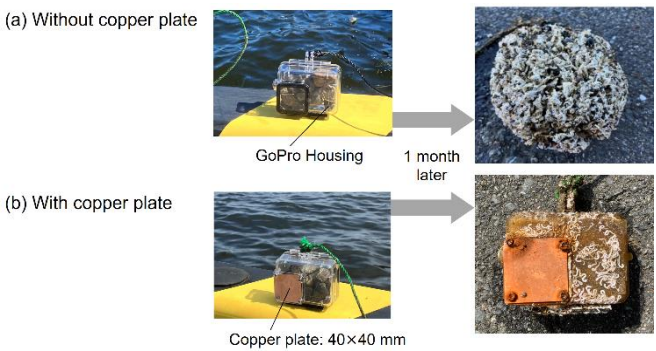

Figure 2: Anti-biofouling effect of copper material.

## III. ANTI-BIOFOULING LENS-LESS CAMERA SYSTEM

In this section, we will show the basic configuration of our proposed camera system. Then, we will also describe the signal processing techniques essential for realizing the lens-less camera utilized in this study.

### A. Design of functionalized plate (photo mask)

First, we conducted a preliminary experiment to determine material of the functionalized plate. By considering the opinions obtained from the conventional works, as described in the Section II-B, we have decided to use copper (Cu) as a base material of the functionalized plate. Fig. 2 shows an example of our preliminary experiments over a month. From Fig. 2(a), we can clearly see that the mockup camera is fully covered by biofouling within one months. On the other hand, Fig. 2(b) suggests that surface of the mockup is maintained thanks to anti-biofouling effect of copper. When microorganisms adhere to the surface of copper, cell membrane gets damaged by generated copper ion, and the microorganism are also affected by ion released from copper.

From a design perspective, if we surround the lens with copper, the crucial optical aperture remains vulnerable. If organisms are to land in the unprotected area and burrow into that weak point, it would be trouble on imaging. Therefore, we have decided that replacing the lens with copper and create numerous micro windows for light intake there.

### B. Prototype of camera unit

Fig. 3 shows the developed camera unit with bio-repellence functionality. This camera unit has also the imaging function that consists of projecting an image on the image screen from the subject outside through the micro patterns as multiple pinholes on the aperture area instead of lens [7], as shown in Fig. 3(a). Fig. 3(b) shows a magnified view of micro patterns which are embedded 2×2 mm area of the copper plate by using the microfabrication process.

Fig. 3(c) shows the cross-section of the camera unit. The size of the holes and the distance between the sensor and the aperture are constrained by the relationship formula for the optimal pinhole size, which will be discussed later. The spacer is mode by 3D printer and it was designed just avoiding the light entering from around. It will be necessary to prevent reflection on the back of the glass. For the imaging sensor, the size of the sensor has limited flexibility due to cost and usability considerations. Therefore, we use a practical image sensor (Basler camera module, Sensor size: 4.32 × 7.68 mm, Pixel pitch: 2 μm) as a camera device.

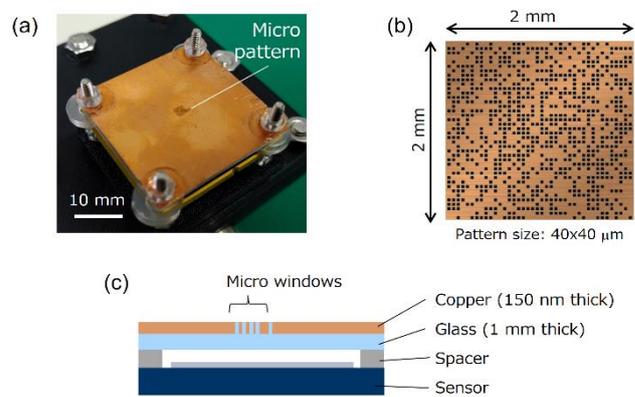

Figure 3: Overview of the developed camera unit.

## C. Basic principle of image reconstruction

In the context of geometric optics, the principle of lens-less imaging using coded apertures can be easily understood by considering it as a multiplexing of pinhole imaging. As shown in Fig. 4, pinhole imaging involves selecting the direction of light propagation through a pinhole to obtain an inverted real image on the image sensor. When the spread of transmitted light on the sensor can be neglected, in the context of geometric optics, the imaging process can be modeled as a normalized similarity transformation from the display image $I(x)$ to the sensor image as

$$f(u) = I(u/z)/z^2. \quad (1)$$

However, since the actual pinhole has a size $a$, the spread of the transmitted light cannot be ignored, and blurring proportional to $a$ is inevitable. When attempting to reduce this blurring by decreasing the pinhole size, diffraction—a wave behavior of light—becomes significant [11]. As a result, blurring inversely proportional to $\lambda a/d$, caused by phase differences within the pinhole, becomes dominant [12]. Due to the trade-off associated with the size of the pinhole, the optimal size is proportional to $\sqrt{\lambda d}$, with an empirical coefficient of approximately 1.9. When $\lambda$ = 500 nm and $d$ = 2 mm, $a$ is approximately 61 $\mu m$, and making it smaller or larger will not improve resolution [13].

Since the size of the pinhole is about one-thousandth of a lens having 2 mm diameter, it leads to a significant insufficiency in the amount of light passing through the aperture. To compensate, approximately 1,000 pinholes are used to increase the light quantity to a level comparable to that of a lens [14]. In this case, the dispersion region of the pinholes should be narrow enough that the light from the same point on the subject can be considered uniform (e.g., for a subject 40 mm away from the aperture, the region would be 2 mm square). The image obtained from the sum of the shifted pinhole images $f(u - a_k)$ centered at points $a_k$ ($k$ = 1…1000) on the aperture area $A$, is a simple explanation of lensless imaging as a sum of shifted pinhole images. Furthermore, by using a transmission distribution $k(a)$ instead of a discrete sum in a relatively dense region, by using a transmission distribution $k(a)$ instead of a discrete sum in a relatively dense region, such as $f(u - a_1) + f(u - a_2) + \cdots$, we obtain:

$$g(u) = \iint_A f(u-a)k(a)da = [f * k](u) \quad (2)$$

which is also known as a convolution integral, where $*$ denotes the convolution of the transparency distribution $k$ on the area $A$ with the image $f$ formed by pinhole imaging.

If the imaging process can be represented by this model, the image restoration can be achieved using deconvolution. Deconvolution can be performed by precomputing the deconvolution filter $r$, and the restored image $h$ can be obtained through the convolution operation $h = g * r$ [15,16]. In practice, the computationally expensive convolution operations $f * k$ and $g * r$ can be simplified into element-wise multiplications using the Fourier transform properties, such as $F[f * k] = F[f] \cdot F[k]$. Similarly, from $F[g * r] = F[g] \cdot F[r]$, the restored image $h$ can be obtained as

$$h(u) = F^{-1}[F[g]/F[k]](u) \quad (3)$$

where $/$ denotes element-wise division.

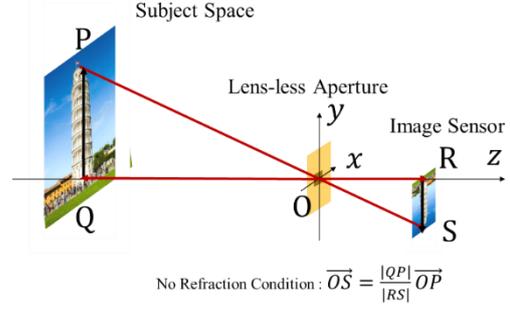

Figure 4: Basic model of pinhole imaging.

However, in practical applications, this restoration operation is often unstable due to errors and noise [15,17]. To address this, and to avoid division by zero while smoothing the characteristics, a parameter $\epsilon$ is introduced, consistent with ridge filtering. This allows for noise-robust restoration, expressed as:

$$h(u) = F^{-1}\left[F[g] \cdot \frac{F[k]^*}{|F[k]|^2 + \epsilon^2}\right](u) \quad (4)$$

## D. Deep-learning based lens-less image reconstruction

The actual process of lens-less imaging is not as simple as described. For instance, there are unavoidable diffraction effects when the aperture is small, interference due to densely packed apertures, multiple reflections, and stray light that hardware cannot eliminate completely, and the signal-to-noise ratio characteristics of the pixels [14,16]. Instead of formalizing these factors into the restoration computation, research is progressing to incorporate them directly into deep learning [18]. The operations for restoration often use an encoder-decoder structure, such as U-Net, which is an advanced form of the autoencoder architecture commonly used in image transformation [19]. The encoder extracts feature from the image, and the decoder reconstructs the image from those features. The encoder-decoder model can be written as:

$$y = \text{dec}(\text{enc}(x, p), q) \quad (5)$$

where $x$ and $y$ are the images from the source domain and target domain, respectively.

Deep learning has mechanisms for learning the convolution operations, which at first glance, might seem well-suited for learning deconvolution operations. However, the convolution kernels in deep learning perform the convolution operation directly, and this does not necessarily lead to high computational efficiency. In the case of lensless imaging, learning global convolution over the entire image using deep learning can be computationally expensive [19,20].

On the other hand, in studies where the global convolution operation is performed not by convolutional layers but by using Transformers for lens-less image restoration, very high-quality reconstructed images have been presented [7]. In a Transformer, the Self-Attention mechanism models relationships between tokens, where a token represents a basic unit of input, such as a word or, in the case of image processing, an image patch [21]. These tokens are generally represented as array variables with shape $(n, d)$, where $n$ represents the number of tokens and $d$ represents the feature size of each token. For each token or patch $p$, the matrix products $pq$, $pk$ and $pv$ are computed, where $q$, $k$, and $v$ are parameter matrices of size $(d, l)$. The attention scores for each token can be calculated using the softmax function to obtain normalized weights, and the final output is written as:

$$r = \text{softmax}\left(\frac{(pq)(pk)^T}{\sqrt{d_k}}\right) pv \qquad (6)$$

where $d_k$ denotes the dimensions of the key features and we can see that the computation models both the token mixing and feature mixing separately, which makes equivalent mixing of tokens having different positions in the token field.

ViT (Vision Transformer) is a Transformer designed for images [23]. Since it requires tokens, the concept of image patches is used as a substitute. Image patches are created as follows:

$$p = \text{reshape}\left(T^*(R^*(x)), (n, d)\right). \qquad (7)$$

Here, $R^*(x)$ reshapes the shape $(H, W, C)$ to $(H/P, P, W/P, P, C)$ and $T^*(x)$ transposes the dimensions to $(0, 2, 1, 3, 4)$, and finally, reshapes it to $(n, d)$, where $n = HWP^{-2}$ is the number of patches and $d_k = CP^2$ is the flattened patch dimension. By treating each patch as a token, ViT processes images as same self-attention as in used for sequences. This allows the model to capture both local details within a patch and global dependencies across the entire image. However, as in the following discussion, it requires $n^2 \sim (HW)^2$ order of the complexity that prevents the practical usage in large scale images [7]. To address this issue, a commonly used method is Axial Attention [21], which has also been adopted in prior studies. In the study, the Axial Attention is used to reduce computational complexity by applying attention separately along the height and width axes:

$$r = \prod_{p,q,k,v=p_i,q_i,k_i,v_i(i\in\{h.w\})} \text{softmax}\left(\frac{(pq)(pk)^T}{\sqrt{CP}}\right) pv \qquad (8)$$

The patches used for Axial Attention are generated using Overlapped Patch Embedding, which applies convolutional layers with kernel size $2P - 1$ and stride $P$ as

$$p = \text{Conv2D}(x, \{\text{stride}: P, \text{kern}: 2P - 1, \text{padd}: P - 1\}). \qquad (9)$$

This method helps the model preserve local continuity while reducing the computational burden.

*E. Proposed reconstruction method by using gMLP (gated Multilayer Perceptron)*

Using Transformers for large-scale images has shown that Axial Attention can effectively reduce computational resources. This is supported by results reported in [21]. However, is this approach truly the best option? Axial Attention processes each axis independently, which might not be ideal for capturing object shapes. As mentioned in [22], "axial processing may not always capture object shapes accurately." In standard 2D convolutions, the vertical and horizontal axes cannot be processed independently because the convolution kernel isn't separable. This means the entire image must be processed together. Therefore, an alternative method is needed to capture global correlations effectively.
In response to this, we propose a gMLP based lens-less image reconstruction, that uses gated multi-layer perceptron gMLP instead of the transformer blocks in the encoder network of the encoder-decoder models.

A typical gMLP contains a single or repetitive set of components called a patch mixer and query mixer spatial unit and a token mixer. The following table shows the inputs, outputs, and parameters of these modules, where $p$ is the input of the gMLP, $q, s, t$ and $r$ are the hidden and the output variables, $u, v$ and $w$ are the parameter matrices. The size $(n, d)$ is the number and the dimension of the token $p$ (or a serialized patch), $h$ is a hidden dimension determined by $md$ where $m$ is a MLP ratio (typically 6), and $l$ is an output dimension determined by network designer.

In overall structure of the proposed method, the GMLP block is integrated into the **encoder network** with the patch embedding unit same as the ViT based (Not the Axial Transformer Based) encoder and the encoder combined with decoder also same as ViT based structure. When $x$ and $y$ area the input and the output of the gMLP, i.e. $x$ is connected to the input of patch-embedding unit and and the output $t$ of the GMLP module is connected to $y$, the single block whole process can be expressed as:

$$r = h\left(\left((f(pw)u + b) \circ f(pw)\right)v\right) \qquad (10)$$

where $\circ$ denotes the element-wise product. In this expression, we can see that the term $(f(pw)u \circ f(pw))v$ enables both the feature mixing and token mixing efficiently similar to ViT.
In other word, the mechanism of the gMLP process the both; axis of images at the same time without the unoptimized process of the axis partition shown in Axial Attentions.

Further, while adhering the constraints of the axis inseparability, the method also achieves the complexity reduction to $o(n)$ as explains with table.1. The table is also useful to estimate the number of the multipliers in (10) and for each module it can be estimated as $ndh$, $nhh$, $nh$, $nhl$, which leads to the total counts as $nh(d + (h + 1) + h + l)$. Which explicitly shows that the complexity order is $o(n)$. Similar derivation provides the estimation of parameters as $h(d + (h + 1) + l)$ whose ratio between them is $n:1$. This number will be also used in latter discussion.

Table1: Inputs-outputs parameters of gMLP modules.

|  | Input | Output | Parameter |
| --- | --- | --- | --- |
| Patch mixer | $p \in \mathbb{R}^{n\times d}$ | $q = f(pw)$ | $w \in \mathbb{R}^{d\times h}$ |
| Spatial Gate | $q \in \mathbb{R}^{n\times h}$ | $s = qu + b$ | $u \in \mathbb{R}^{h\times h}$ |
| Gated mixer | $s \in \mathbb{R}^{n\times h}$ | $t = q \circ s$ | ---------------- |
| Token mixer | $t \in \mathbb{R}^{n\times l}$ | $r = h(tv)$ | $v \in \mathbb{R}^{h\times l}$ |

## IV. EXPEIMENTS

### A. Exmerimental setup and image reconstructon

We constructed a system that actually takes images using a prototype lens-less camera, and demonstrated the quality of the restoration results of lens-less photography. As shown in Fig. 5, the system is roughly composed of a display (5.1 inch, 1920×1080 pixels) that displays photo images like a slideshow, a prototype camera that is positioned to take pictures of the displayed image, a computer that records and distributes the images taken by the prototype camera, and an image conversion module that converts the images using a restoration algorithm. The point of this figure is that it is possible to simultaneously take a picture of what the restored image should look like.

On the other hand, it has the disadvantage that it cannot reflect the actual image if it has been altered by some physical phenomenon. Fig. 6 shows the results of an attempt to restore an image using deep learning as a restoration algorithm. Fig. 6(a) shows the image displayed on the display, Fig. 6(b) shows an image taken by the prototype camera of this research using lens-less photography, and Fig. 6(c) shows the result of restoration using the deep learning algorithm. From this figure, we can see that the combination of lens-less photography and restoration is capable of restoring objects at a level that can be perceived by humans at least.

### B. Comparison of the Computational Costs

To compare the computational costs of the proposed methods with the previous study, the estimations of the required time and memory for three models: ViT-Self-Attention, ViT-Axial-Attention, and gMLP based on (6, 8, 10) respectively.

The basis for this estimation is the number of parameters and the total number of floating-point operations (FLOPs). Instead of counting all the FLOPs, we approximate it by the number of multiplications. When the input shape is (B, H, W, C), where B denotes the batch size and the output size(H,W,L), the formulas for the number of parameters and the FLOPs for each model are as shown in Table 2. These calculations for the number of parameters and multiplications can be derived by dividing the operations between input variables and intermediate variables, and intermediate variables and output variables. With the result of number off the parameters and multiplications, we can consider the required time and required memory. The required time can be estimated based on the FLOPs and the machine's processing speed in terms of FLOPs per second (TFLOPs). The required memory is estimated as at maximum the sum of the number of parameters, the number of the inputs in the addition operations and the multiplication operations. In deep learning, it is typical to reuse memory for addition operations with memory allocated for multiplication operations. Thus, the required memory can be approximated as (Number of parameters + 2 Number of FLOPs) × Precision, where the precision = 4 bytes or 2 bytes.

Based on this, we can predict the actual computation time and the required memory. Assuming we are using a GeForce RTX 4080 with a memory capacity of 15GB and processing power of 45TFLOPs. For ViTs and gMLP, we assume both a single block with embedding of 512and 4 stacked blocks

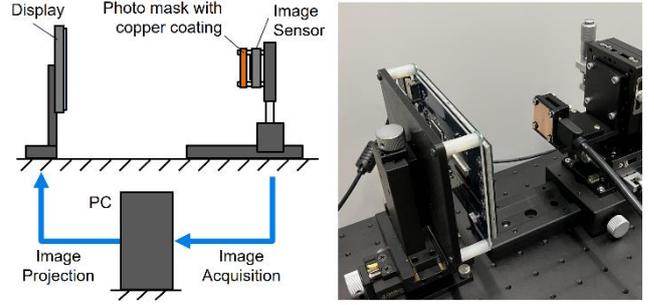

Figure 5: Experimental setup.

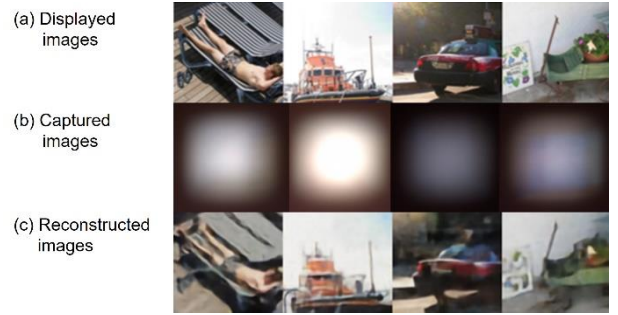

Figure 6: Obtained results by deep-learning image reconstruction.

Table2: Fundamental values for the ViTs and the gMLP.

| Model | #Params | #Multiplications |
|---|---|---|
| ViT (SA) | $2CP^2 \cdot CP^2 +$ $CP^2 \cdot L$ | $2HWC^2 +$ $2(HW)^2 CL$ |
| ViT (AA) | $(H^2 + W^2) \cdot (CP)^2 +$ $2(H + W) \cdot CP \cdot L$ | $2(H + W)C^2 +$ $2(H^2 + W^2)CL$ |
| gMLP | $mCP^2 \cdot CP^2 +$ $mCP^2 \cdot (mCP^2 + 1) +$ $mCP^2 \cdot L$ | $mHWC \cdot CP^2 +$ $mHWC \cdot (2mCP^2 + 1) +$ $mHWC \cdot L$ |

Table3: Estimated properties for the ViTs and the gMLP.

| Image | Model | Embed | Times[s] | VRAM[GB] FP32 | VRAM[GB] FP16 |
|---|---|---|---|---|---|
| LargeIcon (160,160) | ViT-S | (64,128,256,512), | 3.19E+02 | 1.15E+05 | 5.74E+04 |
| | VIT-A | (64,128,256,512), | **0.025** | **8.98** | **4.49** |
| | gMLP | (64,128,256,512), | 0.282 | 1.02E+02 | 5.08E+01 |
| | ViT-S | (512), | 4.47E+01 | 1.61E+04 | 8.05E+03 |
| | VIT-A | (512), | **0.003** | **1.26** | **6.30E-01** |
| | gMLP | (512), | 0.012 | 4.19 | 2.10E+00 |
| ImageNet (224,224) | VIT-A | (64,128,256,512), | 4.88E-02 | 1.76E+01 | **8.80** |
| | gMLP | (64,128,256,512), | 0.554 | 1.99E+02 | 9.96E+01 |
| | VIT-A | (512), | **0.007** | **2.47** | **1.23** |
| | gMLP | (512), | 0.023 | 8.22 | 4.11 |
| VGA (800,640) | VIT-A | (512), | **0.072** | 2.58E+01 | **12.91** |
| | gMLP | (512), | 0.233 | 8.38E+01 | 4.19E+01 |
| HiVision (1920,1080) | VIT-A | (512), | **0.342** | 1.23E+02 | 6.16E+01 |
| | gMLP | (512), | 0.963 | 3.47E+02 | 1.73E+02 |

cases for each with embedding dimensions of 64, 128, 256,and 512, with patch size both (4, 4). For gMLP, we assume an MLP Ratio of 6 for all the cases. Using these assumptions, we calculate the predicted time and required memory for input sizes of HiVision (1920×1080), VGA (640×480), ImageNet (224× 224) and Larger Icon (160×160) images.

The results are shown in Table 3. In this table, bold digits shows that the value is lower than 1 second or 15GByte, meaning practical level necessarily. From the table, we can see that even for Icon size, VRAM in GPU is insufficient for ViT-S (Self attention), thus it is not shown in larger images. For HiVision size, all models face challenges in memory constraints. Between gMLP and ViT-A (Axial Attention), the required memory for gMLP tends to be about 5 to 10 times larger than that of ViT-A. To compare with these with fair criteria in terms of memory, the combination of 4 blocks ViT-A and single block gMLP for the larger icon size images are best choice among in the available configurations.

*C. Performance Comparison*

We conducted an experiment to evaluate whether images displayed on a screen could be reconstructed using a prototype lens-less camera. In the experiment, we used the lens-less camera equipped with the coded aperture explained in Fig. 3 and projected images onto the display shown in Fig. 5, simulating a virtual object. These results were collected as a dataset for training and evaluation. The distance between the display and the aperture was set to 18 cm, and the distance between the aperture and the sensor was set to 2 mm.

For the training and evaluation datasets, we prepared 10k and 50k images, respectively, for supervised reference, using the captured lens-less images to query and match the corresponding reference images. These reference images were randomly selected from the ILSVRC ImageNet 2012 dataset, which is a well-known benchmark for image classification tasks. Specifically, we randomly selected 100 images and 500 images from 100 randomly chosen categories within the dataset. Since the images had varying sizes, we performed center cropping and resizing to ensure that all images were processed into a consistent square format, suitable for the model.

We adopted PyTorch framework for model designing and training. Timm modules were used to assist designing the gated MLP model inside the encoder. As explained previously, the encoder contains a single block of a gated MLP with embedding size of 512 and patch size (4, 4). For the reference model ViT-A (axial attention) we select 4 stacked axial attention blocks structure with the embedding size off (64, 128, 256, 512), patch size (4, 4) following to [7]. For both models, batch size 4 and a batch normalization are applied for each block.

As the decoder network, that generates the deconvoluted clear images from the encoder's outputs, the same structure with 4 stacked fully convolution with kernel size (3,3) is applied for both proposed and reference models. The last layer is connected to a resize layer which provide the same size images as that of the reference images. The resized output and the reference image are input to cost calculation layer and the errors to the reference are back propagated to the input variables and the trainable parameter for each layer. For the cost function. Mean squared error (MSE) is assigned to measure the prediction error. The optimization by the backpropagation is brought by one of the refined gradient descent methods, ADAM-W, with learning rate of 6E-5, and weight decay of 0.1. We also applied warmup scheduling for the beginning several epochs to get stability in the optimization.

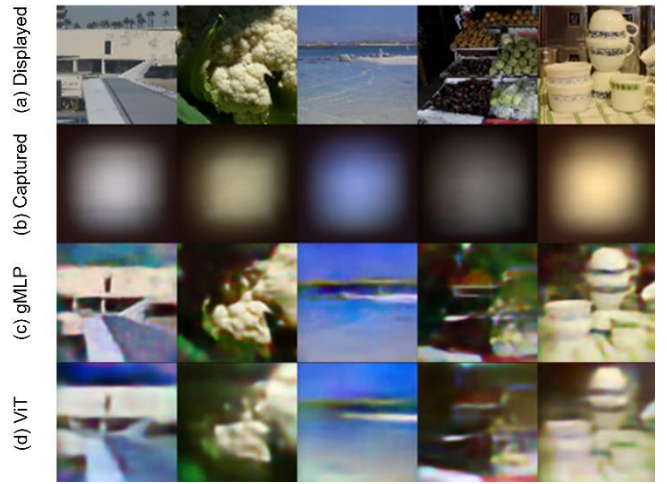

Figure 7: Comparison between computational image reconstruction methods.

**Table 4: Statistical analysis of obtained results.**

| Model | In-Out size input->output | Param size [MB] | Train size | PSNR [dB] /SSIM |
|---|---|---|---|---|
| gMLP | 160-80 | 203.08 | 59,000 | 19.80/0.6149 |
|  | 160-80 | 203.08 | 9,000 | 18.48/0.5254 |
| ViT Axial | 160-80 | 200.4 | 59,000 | 17.67/0.4761 |
|  | 160-80 | 200.4 | 9,000 | 17.30/0.4279 |

An NVIDIA RTX 4080 GPU with 15 GB of VRAM was used for computational power and memory, with an Intel i7-9700K CPU 3.0GHz with 32 GB Ram for efficient data preprocessing. The system ran on Ubuntu 20.04 LTS, and it takes about 10 hours for 50 epoch iteration for training size of 9000 images and 50 hours for 50 epoch with 49000 images. After the training processes, we input the images for evaluation, that is not used in training phase, and obtain clear images from the lens-less completely blurred image.

Fig. 7 shows the experimental results where the evaluation sample set of the displayed images (Fig. 7(a)), the captured images (Fig. 7(b)), the recovered by the proposed method (gMLP, Fig. 7(c)), and the reference method (ViT-A, Fig. 7(d)). The reconstructor were both trained with 49,000 (larger) datasets. The obtained reconstructed result is reasonable that the completely blurred lens-less image has been keeping the information for their resurrection data. Comparison of gMLP and ViT-A images show that the proposed gMLP based method supplied slightly clearer and sharper images than that of the reference methods.

Table 4 shows the statistical evaluation of 1000 images. The table has a column for averaged scores of pSNR and SSIM, typical indices to evaluate quality of a reconstructed image with larger scores meaning better reconstruction. Both values increase in the rows for the gMLP based model indicating the superiority of the proposed method in the image reconstruction. One might be interested in how the difference behaves in terms of pixel intensity. PSNR is defined as $10\log_{10}(MAX^2/MSE)$, where MAX represents the maximum possible pixel value. with can estimate the rate of the intensity error rate. Let as assume $p$ and $q = p(1 + \rho\varepsilon)$ as pixel intensities at pixel $x$ in a displayed image and its reconstructed image, where $\rho$ is a constant value expressing intensity error rate and $\varepsilon$ is a standard normal distribution sample at each pixel. Since $p$ and $\varepsilon$ are sampled independently, pSNR is $10 \log_{10}(\text{MAX}^2 \rho^2 \mathbb{E}[p(x)^2])$ and we can assume MAX and $\mathbb{E}[p(x)^2]$ is same for both the proposed and reference method, thus the rate of the intensity error rate can be calculated as rate = power(10, dp/2) where dp denotes the difference of pSNR between the proposed and the reference method. SSIM stands for Structural Similarity Index, which evaluates the overall similarity between images by combining structural information, luminance, and contrast, whereas pSNR is focused solely on luminance. SSIM does not separate these components but provides a comprehensive assessment of the images. Unlike pSNR, SSIM is normalized and easier to interpret as a score. Intuitively, for example, if other factors remain equal, SSIM can reflect the area overlap between two images. The difference between SSIM values of 0.615 and 0.47 corresponds to a 25% discrepancy in area overlap, which represents a relatively noticeable difference.

## V. DISCUSSIONS

### A. Performance of imaging reconstruction

In the context of geometric optics, lens-less imaging is modeled as a convolution of the through a pinhole. When performing lens-less deconvolution, the process is similar to reconstructing images captured through a pinhole. The deconvolution can be implemented using FFT (Fast Fourier Transform), though it is vulnerable to errors and noise. To address this, deep learning incorporates deconvolution into the learning process, enabling the entire imaging task to be learned.

However, convolutional layers in deep learning are computationally expensive for learning global convolutions. Transformers, on the other hand, offer more efficient control as they are not dependent on spatial distances. Despite this, ViT's self-attention mechanism tends to be computationally expensive and lacks long-term efficiency. The axial attention can be used to reduce computational costs to linear complexity while preserving essential image features. The Axial attention, however, encounters difficulties when modeling global convolutions due to the indivisibility of the axes (i.e., vertical and horizontal axes cannot be computed separately). gMLP solves this problem by addressing both the indivisibility and the computational cost simultaneously. Finally, experiments with real lens-less imaging hardware have demonstrated that gMLP is stable and delivers reasonable performance, especially for large-scale global convolution operations.

In the future work, design parameters of the coded aperture and the image reconstruction algorism should be mutually tuned and optimized to improve the quality of the reconstructed images and the lifetime of the observation system.

### B. Experiments at actual site

The effectiveness of copper material in preventing marine organisms from fouling was shown in the Section III-A, and the quality of lens-less image reproduction was shown in the Section IV. As described in the Section III, the main question is whether it is possible to achieve both the imaging and anti-fouling (bio-repellence) at the same time.

Therefore, we have installed the developed camera system in an aquarium in a coastal environment where seawater can be taken in and circulated, as shown in Fig. 8 and created an environment in which it could take the same images in front of it. We have built a wireless monitoring system that can automatically capture the images in every 30 sec and left it for about 30 days.

As a result, copper deposition with a thickness of less than 150 nm was peeled off in seawater within a couple of days, as shown in Fig. 8(c). Lens-less imaging becomes completely ineffective when the encoded pattern is lost. To prevent or delay the disappearance of thin copper films in seawater, both chemical and physical measures are available. While chemical-based approaches will be addressed separately, the most effective physical measure is likely to increase the thickness of the copper. In fact, as shown in Fig. 2, the copper plate approximately 100 microns thick was submerged in seawater and maintained their effectiveness without disappearing. From the perspective of the antifouling function, the thicker material would be better. This trend is common across many other physical functions such as pressure resistance, heat resistance, light shielding, electromagnetic shielding, infrared and ultraviolet shielding, X-ray shielding, and radiation shielding, and so on.

### C. Balace between image reconstruction and anti-fouling

How does the guideline that "thicker is better" hold up in the context of image restoration, specifically in imaging? In fact, it has been observed that when the thickness exceeds a certain level, the field of view narrows, similar to a pinhole, and the image quality deteriorates as a result. Therefore, in the image reconstruction, it is required to find a reasonable balance between the thickness of the mask (hole) and the field of view of each pinhole. Since one of major objective of this study is to achieve both clear reconstruction in lens-less imaging and the integration of non-image-related effects brought about by the aperture material or microstructure, we will try to find good balance to establish a highly durable camera system.

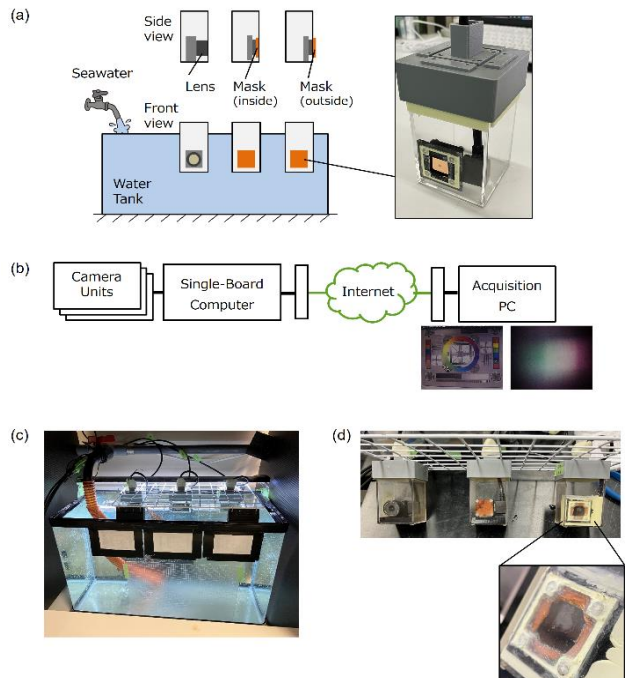

Figure 8: The developed camera system for on-site experiment.

## VI. Summary

In this paper, we showed the overview of the proposed approach which is actively combined with the functionalized lens-less camera and the image reconstruction computing algorithm to realize a long-life in-situ ocean monitoring. The basic working principle of the image reconstruction approach was demonstrated by using the lens-less camera with the Cu-based coded aperture with micro patterns. The main results of this paper are as follows:

(1) We examined the compatibility of two functions—light intake for imaging and biofouling prevention (referred to here as bio-repellent functionality)—in a lens-less imaging system using a coded aperture with antifouling properties.

(2) In doing so, we demonstrated that the still-developing potential of lens-less cameras is capable of restoring images with VGA-level resolution. Additionally, we implemented an original method that addresses mathematical challenges in existing techniques, showing that it can improve performance in actual devices.

(3) Furthermore, we found that thickness is a key factor in balancing the imaging functionality of lens-less cameras with other non-imaging functions applied to the aperture. Specifically, when the thickness is less than the equivalent pinhole diameter in pinhole imaging, it is highly feasible to achieve both functions.

In the future works, temporal variation in the image reconstruction performance will be examined in more detail under ocean simulated environment. Especially, it is important to confirm limitation of durability and antifouling performance of the copper-based mask. The design parameters of the coded aperture and the image reconstruction algorism will be optimized and examined for realizing a new type of camera system to inspect the undersea world.

## Acknowledgement

The authors would like to acknowledge Dr. Kogiku Shiba and Mr. Jiro Takano at the Shimoda Marine Research Center, University of Tsukuba for their support in conducting experiments. The authors thank Ms. Emiko Sekimoto at the Advanced Research Laboratory, Sony Group Corporation for her helpful support into this work.